\documentstyle[prb,aps,amsfonts,amssymb,amsmath,twocolumn,epsfig]{revtex}
\begin{document}
\draft

\wideabs{

\title{Spin injection and spin accumulation in permalloy-copper mesoscopic spin valves.}

\author{F.J. Jedema, M.S.Nijboer, A.T. Filip, B.J. van Wees}
\address{Department of Applied Physics and Materials Science Center, University
of Groningen,\\ Nijenborg 4, 9747 AG Groningen, The Netherlands}

\date{\today}

\maketitle
\begin{abstract}
We study the electrical injection and detection of spin currents in a lateral
spin valve device, using permalloy (Py) as ferromagnetic injecting and
detecting electrodes and copper (Cu) as non-magnetic metal. Our multi-terminal
geometry allows us to experimentally distinguish different magneto resistance
signals, being 1) the spin valve effect, 2) the anomalous magneto resistance
(AMR) effect and 3) Hall effects. We find that the AMR contribution of the Py
contacts can be much bigger than the amplitude of the spin valve effect, making
it impossible to observe the spin valve effect in a 'conventional' measurement
geometry. However, these 'contact' magneto resistance signals can be used to
monitor the magnetization reversal process, making it possible to determine the
magnetic switching fields of the Py contacts of the spin valve device. In a
'non local' spin valve measurement we are able to completely isolate the spin
valve signal and observe clear spin accumulation signals at $T=~4.2$ K as well
as at room temperature. We obtain spin diffusion lengths in copper of $1$
micrometer and $350$ nm at $T=4.2$ K and room temperature respectively.

\end{abstract}}
\section{introduction}
Spintronics is a rapidly emerging field in which one tries to study or make
explicit use of the spin degree of freedom of the electron. Sofar, the most
well known examples of spintronics are the tunneling magneto resistance (TMR)
of magnetic tunnel junctions, and the giant magneto resistance (GMR) of
multilayers\cite{tedrow,gijs1,ans1}. A new direction is emerging, where one
actually wants to inject spin currents, transfer and manipulate the spin
information, and detect the resulting spin polarization. Because of spin-orbit
interaction, the electron spin can be flipped and consequently a spin polarized
current will have a finite lifetime. For this reason it is necessary to study
spin transport in systems, where the 'time of flight' of the electrons between
the injector and detector is shorter than the spin flip time. In diffusive
metallic systems, this corresponds to typical length scales of a micrometer. We
use a lateral mesosopic spin valve, to access and probe this length
scale\cite{jedema}. It consists of a ferromagnetic injector electrode and
detector electrode, separated over a distance $L$ by a normal metal region, see
Fig. \ref{sample}.

In this paper a review of the basic model for spin transport in the diffusive
transport regime will be given and applied to our multi-terminal device
geometry. Secondly, a description and measurements of the magnetic switching
behavior of the Py electrodes used in the spin valve device will be presented.
Finally measurements of the spin valve effect in a 'conventional' and
'non-local' geometry will be shown and analyzed using the model for spin
transport in the diffusive regime.

\section{Theory of spin injection and accumulation}
We focus on the diffusive transport regime, which applies when the mean free
path $l_e$ is shorter than the device dimensions. The description of electrical
transport in a ferromagnet in terms of a two-current (spin-up and spin-down)
model dates back to Fert and Campbell \cite{fert1}. Van Son et al. \cite{son1}
have extended the model to describe transport through ferromagnet-normal metal
interfaces. A firm theoretical underpinning, based on a Boltzmann transport
equation has been given by Valet and Fert \cite{valet1}. They have applied the
model to describe the effects of spin accumulation and spin dependent
scattering on the giant magneto resistance (GMR) effect in magnetic
multilayers. This "standard" model allows for a detailed quantitative analysis
of the experimental results.

An alternative model, based on thermodynamic considerations, has been put
forward and applied by Johnson \cite{john1}. In principle both models describe
the same physics, and should therefore be equivalent. However, the Johnson
model has a drawback in that it does not allow a direct calculation of the spin
polarization of the current ($\eta$ in refs.\ref{john1} and \ref{john2}),
whereas in the standard model all measurable quantities can be directly related
to the parameters of the experimental system.

The transport in a ferromagnet is described by spin dependent conductivities:

\begin{eqnarray}
\sigma_\uparrow & = & N_\uparrow e^2 D_\uparrow, \;\textrm{with } D_\uparrow =
\frac{1}{3} v_{F\uparrow} l_{e\uparrow}\
\label{conductivityup}\\
\sigma_\downarrow & = & N_\downarrow e^2 D_\downarrow,\;\textrm{with }
D_\downarrow = \frac{1}{3} v_{F\downarrow} l_{e\downarrow}\;,
\label{conductivitydown}
\end{eqnarray}

where $N_{\uparrow,\downarrow}$ denotes the spin dependent density of states
(DOS) at the Fermi energy, and $D_{\uparrow,\downarrow}$ the spin dependent
diffusion constants, expressed in the spin dependent Fermi velocities
$v_{F\uparrow,\downarrow}$, and electron mean free paths $l_{e
\uparrow,\downarrow}$. Note that the spin dependence of the conductivities is
determined by \emph{both} density of states and diffusion constants. This
should be contrasted with magnetic F/I/F or F/I/N tunnel junctions, where the
spin polarization of the tunneling electrons is determined by the
spin-dependent DOS. Also in a typical ferromagnet several bands (which
generally have different spin dependent density of states) contribute to the
transport. However, provided that the elastic scattering time and the interband
scattering times are shorter than the spin flip times (which is usually the
case) the transport can still be described in terms of well defined spin up and
spin down conductivities.

Because the spin up and spin down conductivities are different, the current in
the bulk ferromagnet will be distributed accordingly over the two spin
channels:

\begin{eqnarray}
j_\uparrow & = & (\frac{\sigma_{\uparrow}}{e})\frac
{\partial\mu_{\uparrow}}{\partial x}
\label{currentup}\\
j_\downarrow & = & (\frac{\sigma_{\downarrow}}{e})\frac
{\partial\mu_{\downarrow}}{\partial x}\;, \label{currentdown}
\end{eqnarray}

where $j_{\uparrow\downarrow}$ are the spin up and spin down current densities
and $e$ is the absolute value of the electronic charge. According to eqs.
\ref{currentup} and \ref{currentdown}  the current flowing in a bulk
ferromagnet is spin polarized, with a polarization given by:

\begin{equation}
\alpha_F=\frac{\sigma_\uparrow-\sigma_\downarrow}{\sigma_\uparrow+\sigma_\downarrow}.
\label{polarization}
\end{equation}

The next step is the introduction of spin flip processes, described by a spin
flip time $\tau_{\uparrow\downarrow}$ for the average time to flip an up-spin
to a down-spin, and $\tau_{\downarrow\uparrow}$ for the reverse process. The
detailed balance principle imposes that
$N_\uparrow/\tau_{\uparrow\downarrow}=N_\downarrow/\tau_{\downarrow\uparrow}$,
so that in equilibrium no net spin scattering takes place. As pointed out
already, usually these spin-flip times are larger than the momentum scattering
time $\tau_e=l_e/v_F$. The transport can then be described in terms of the
parallel diffusion of the two spin species, where the densities are controlled
by spin-flip processes. It should be noted however that in particular in
ferromagnets (e.g. permalloy\cite{dub1}) the spin flip times may become
comparable to the momentum scattering time. In this case an (additional)
spin-mixing resistance arises \cite{gijs1}, which we will not discuss further
here.

The effect of the spin-flip processes can now be described by the following
equation (assuming diffusion in one dimension only):
\begin{equation}
D\frac{\partial^2 (\mu_\uparrow-\mu_\downarrow)}{\partial
x^2}=\frac{(\mu_\uparrow-\mu_\downarrow)}{\tau_{sf}}, \label{diffusion}
\end{equation}

where $D=D_\uparrow D_\downarrow (N_\uparrow+N_\downarrow) /(N_\uparrow
D_\uparrow + N_\downarrow D_\downarrow)$ is the spin averaged diffusion
constant, and the spin-relaxation time $\tau_{sf}$ is given by: $1/\tau_{sf}=
1/\tau_{\uparrow\downarrow} + 1/\tau_{\downarrow\uparrow}$. Using the
requirement of current conservation, the general solution of eq.
\ref{diffusion} for a uniform ferromagnet or non-magnetic wire is now given by:

\begin{eqnarray}
\mu_\uparrow & = & A+Bx+\frac{C}{\sigma_\uparrow}
exp(-x/\lambda_{sf})+\frac{D}{\sigma_\uparrow} exp(x/\lambda_{sf})
\label{solutionup} \\
\mu_\downarrow & = & A+Bx-\frac{C}{\sigma_\downarrow}
exp(-x/\lambda_{sf})-\frac{D}{\sigma_\downarrow} exp(x/\lambda_{sf})\;,
\label{solutiondown}
\end{eqnarray}

where we have introduced the spin flip diffusion length
$\lambda_{sf}=\sqrt{D\tau_{sf}}$. The coefficients A,B,C, and D are determined
by the boundary conditions imposed at the junctions where the wires is coupled
to other wires. In the absence of spin flip scattering at the interfaces the
boundary conditions are: 1) continuity of $\mu_\uparrow$, $\mu_\downarrow$ at
the interface, and 2) conservation of spin-up and spin-down currents
$j_\uparrow$, $j_\downarrow$ across the interface.

\section{Spin accumulation in multi-terminal spin valve
structures} 
\begin{figure}[tbh]
\centerline{\psfig{figure=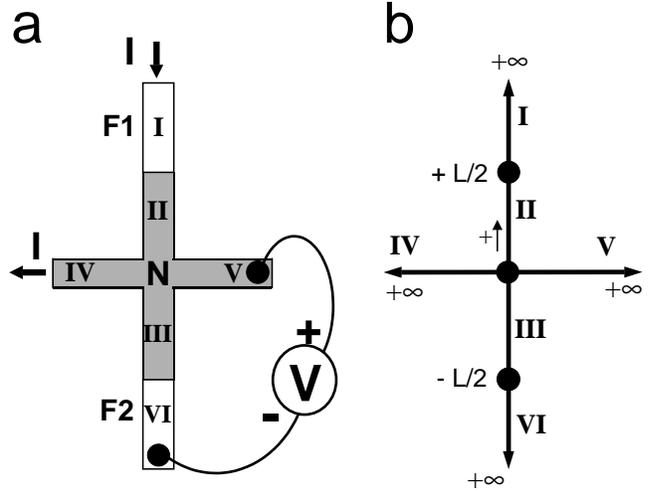,width=86mm}} \caption{(a) Schematic
representation of the multi-terminal spin valve device. Regions \emph{I} and
\emph{VI} denote the injecting ($F_1$) and detecting ($F_2$) ferromagnetic
contacts, whereas regions \emph{II} to \emph{V} denote the four arms of a
normal metal cross ($N$) placed in between the two ferromagnets. A spin
polarized current is injected from region \emph{I} into region \emph{II} and
extracted at region \emph{IV}. (b) Diagram of the electrochemical potential
solutions (eqs. \ref{solutionup} and \ref {solutiondown}) in each of the six
regions of the multi-terminal spin valve. The nodes represent the origins of
the coordinate axis in the 6 regions, the arrows indicate the (chosen)
direction of the positive x-coordinate. Regions \emph{II} and \emph{III} have a
finite length of half the separation distance between the Py electrodes: $L/2$.
The other regions are semi-infinite.} \label{diagram}
\end{figure}

We will now apply the "standard" model of spin injection to a
multi-terminal geometry, which reflects our device geometry used in the
experiment, see Fig.\ref{diagram}a.

In our (1-dimensional) geometry we can identify 6 different regions for which
eqs. \ref{solutionup} and \ref{solutiondown} have to be solved according to
their boundary conditions at the interface. The geometry is schematically shown
in Fig.\ref{diagram}b, where the 6 different regions are marked with roman
letters I to VI. According to eq. \ref{solutionup} the equations for the spin
up electrochemical potentials in these regions, assuming parallel magnetization
of the ferromagnetic regions, read:

\begin{align*}
\mu_\uparrow
&=~A-\frac{je}{\sigma_F}x+\frac{2C}{\sigma_{F}(1+\alpha_{F})}exp(-x/\lambda_{F})
\tag{$I$}
\\
\mu_\uparrow
&=~\frac{-je}{\sigma_N}x+\frac{2E}{\sigma_{N}}exp(-x/\lambda_{N})+\frac{2F}{\sigma_{N}}
exp(x/\lambda_{N}) \tag{$II$}
\\
\mu_\uparrow &=~\frac{2H}{\sigma_{N}}exp(-x/\lambda_{N})+\frac{2K}{\sigma_{N}}
exp(x/\lambda_{N}) \tag{$III$}
\\
\mu_\uparrow &=~\frac{je}{\sigma_N}x+\frac{2G}{\sigma_{N}}exp(-x/\lambda_{N})
\tag{$IV$}
\\
\mu_\uparrow &=~\frac{2G}{\sigma_{N}}exp(-x/\lambda_{N}) \tag{$V$}
\\
\mu_\uparrow &=~B+\frac{2D}{\sigma_{F}(1+\alpha_{F})}exp(-x/\lambda_{F})\;,
\tag{$VI$}
\end{align*}

where we have written $\sigma_\uparrow=\sigma_F(1+\alpha_F)/2$ and
$A,B,C,D,E,F,G,H$ and $K$ are 9 unknown constants. The equations for the spin
down electrochemical potential in the six regions of fig. \ref{diagram} can be
found by putting a minus sign in front of the constants $C,D,E,F,H,K,G$ and
$\alpha_F$ in eqs. $I$ to $VI$. Constant $B$ is the most valuable to extract
from this set of equations, for it gives the difference between the voltage
measured with a normal metal probe at the center of the normal metal cross in
fig. \ref{diagram}a and a ferromagnetic voltage probe. Solving the eqs. $I$ to
$VI$ by taking the continuity of the spin up and spin down electrochemical
potentials and the conservation of spin up and spin down currents at the 3
nodes of Fig. \ref{diagram}b, one obtains:

\begin{equation}
B=-je\frac{\alpha_F^2\frac{\lambda_N}{\sigma_N}e^{-L/2\lambda_N}}
{2(M+1)[Msinh(L/2\lambda_N)+cosh(L/2\lambda_N)]}\;, \label{constantb}
\end{equation}
where $M=(\sigma_F\lambda_N/\sigma_N\lambda_F)(1-\alpha_F^2)$.

In the situation where the ferromagnets have an anti-parallel magnetization
alignment, the constant $B$ of eq. \ref{constantb} gets a minus sign in front .
Upon changing from parallel to anti-parallel magnetization configuration (a
spin valve measurement) a difference of $\Delta\mu=~2B$ will be detected in
electrochemical potential between a normal metal and ferromagnetic voltage
probe. This leads to the definition of the so-called spin-coupled or
spin-dependent resistance of $\Delta R=~\frac{2B}{-ejA}$:

\begin{equation}
\Delta R = \frac{\alpha_F^2\frac{\lambda_N}{\sigma_NA}e^{-L/2\lambda_N}}{(M+1)
[Msinh(L/2\lambda_N)+cosh(L/2\lambda_N)]}\;. \label{Rspinfull}
\end{equation}

Equation \ref{Rspinfull} shows that for $\lambda_N << L$, the magnitude of the
spin signal $\Delta R$ will decay exponentially as a function of L. In the
opposite limit, $\lambda_F << L << \lambda_N$, the spin signal $\Delta R$ has a
1/L dependence:
\begin{equation}
\Delta R = \frac{2\alpha_F^2\lambda_N^2}{M(M+1)\sigma_NAL}\;.
 \label{Rspin}
\end{equation}
Actually, for eq. \ref{Rspin} to hold a more precise constraint has to be full
filled, requiring the relation $ML/2\lambda_N>>1$ to be satisfied. However, the
important point to notice is that eq. \ref{Rspin} clearly shows that even in
the situation when there are no spin flip processes in the normal metal
($\lambda_N=\infty$), the spin signal $\Delta R$ is reduced with increasing
$L$. The reason is that the \emph{spin dependent} resistance of the injecting
and detecting ferromagnets remains constant for the two spin channels, whereas
the \emph{spin independent} resistance of the normal metal increases linearly
with L.

Finally, the current polarization \emph{at the interface} of the current
injecting contact, defined as
$P=~\frac{j_{\uparrow}^{int}-j_{\downarrow}^{int}}{j_{\uparrow}^{int}+j_{\downarrow}^{int}}$,
can be calculated. For parallel magnetized ferromagnetic electrodes the
polarization P yields:

\begin{equation}
P=~\alpha_F\frac{Me^{L/2\lambda_N}+2cosh(L/2\lambda_N)}{2(M+1)
[Msinh(L/2\lambda_N)+cosh(L/2\lambda_N)]}\;. \label{polint1}
\end{equation}

In the limit that $L=\infty$ we obtain the polarization of the current at a
single F/N interface:\cite{son1,filip}

\begin{equation}
P=~\frac{\alpha_F}{M+1}\;. \label{polint2}
\end{equation}

\section{sample fabrication}

\begin{figure}[b]
\centerline{\psfig{figure=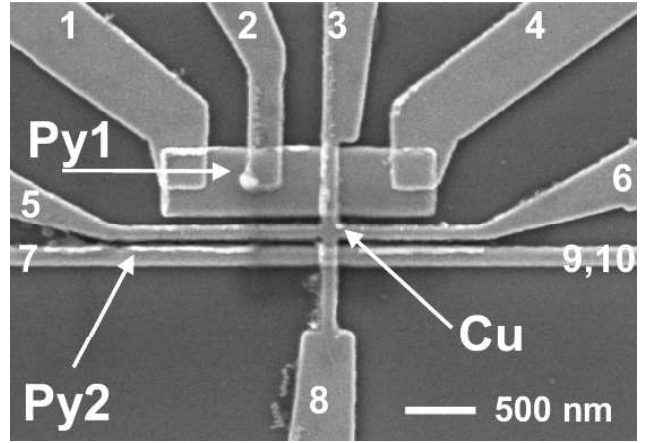,width=86mm}} \caption{Scanning electron
microscope (SEM) picture of the lateral mesoscopic spin valve device. The two
horizontal strips are the ferromagnetic electrodes Py1 and Py2. A copper cross
is placed in between the Py electrodes, which vertical arms lay on top of the
Py electrodes. A total of 10 contacts (not all visible) are connected to the
device.} \label{sample}
\end{figure}

We use permalloy $Ni_{80}Fe_{20}$ (Py) electrodes to drive a spin polarized
current into copper (Cu) crossed strips, see fig. \ref{sample}.

The devices are fabricated in two steps on a thermally oxidized Si wafer by
means of conventional e-beam lithography with PMMA resist. To avoid magnetic
fringe fields from the ferromagnetic electrodes, the $40$ nm thick Py
electrodes were sputter deposited first on a thermally oxidized silicon
substrate, using a $2$ nm tantalum (Ta) adhesion layer and applying a small
B-field of $~3$ mT along the long axis of the Py electrodes. In the second
fabrication step, 50 nm thick crossed Cu strips were deposited by e-gun
evaporation in a $10^{-8}$ mbar vacuum. Prior to the Cu deposition, around $5$
nm of Py material was removed of the Py electrodes by ion milling, thereby
removing the oxide to ensure transparent contacts. The conductivities of the Py
and Cu films were determined to be $\sigma_{Py}=~6.6\cdot 10^6 ~ \Omega ^{-1}
m^{-1}$ and $\sigma_{Cu}=~3.5\cdot 10^7 ~ \Omega ^{-1} m^{-1}$ at RT. At 4.2 K
both conductivities increased by a factor 2.

\section{Magnetic switching of the Py electrodes}

The resistance of a single ferromagnetic strip is a few percent smaller when
the magnetization direction is perpendicular to the current direction as
compared to a parallel alignment. This effect is known as the anomalous magneto
resistance(AMR) effect\cite{coehoorn}. The AMR effect can therefore be used to
monitor the magnetization reversal or 'switching' behavior of the ferromagnetic
Py electrodes\cite{gior1}. Different models can be considered for describing
the magnetization reversal processes in mesoscopic wires\cite{schabes}.

\subsection{Magnetization reversal models}

The simplest description is provided by the Stoner and Wohlfarth model
(SW)\cite{sw}. It assumes a single ferromagnetic domain and coherent
magnetization rotation. Neglecting the magneto-crystalline anisotropy, the
total energy for an ellipsoid of revolution is written as a sum of
magnetostatic and shape anisotropy energies:

\begin{equation}
E=\frac{\mu_{0}M_s^2}{2}(D_z-D_x)\cos^2(\phi-\theta)-\mu_{0}HM_s cos\phi \;,
\label {SW}
\end{equation}
where $M_s$ is the saturation magnetization, $D_z$ and $D_x$ are the
demagnetization factors, and $\phi$ and $\theta$ are the angles between the
magnetization direction and the applied f\mbox{}ield, and, respectively, the
external f\mbox{}ield and the easy axis. The first term on the right of eq.
\ref{SW} represents the shape anistropy energy of the ellipsoid, which is equal
to the magnetostatic self-energy of the particle. For an elongated ellipsoid
along the z-axis the demagnetization factors would be $D_x=0.5$ and $D_z=0$.
The angle between the magnetization and the applied f\mbox{}ield for a given
f\mbox{}ield can be determined analytically by minimizing the total energy. The
switching or coercive f\mbox{}ield as a function of the direction of the
applied f\mbox{}ield reads:

\begin{equation}
H_{c}(\theta)=H_{0}^{sw}(\sin ^{2/3}\theta +\cos ^{2/3}\theta )^{-3/2}\;,
\end{equation}
where $H_{0}^{sw}=M_s(D_z-D_x)$ is the saturation f\mbox{}ield in perpendicular
$(\theta =90^{\circ })$ direction, which corresponds to the demagnetization
field along the (short) x-axis of the ellipsoid of revolution. We thus obtain
an upper value estimate of the switching field for a permalloy ellipsoid of:
$\mu_0 H_{0}^{sw}=~540 mT$, using $M_s=~860 kA/m$.

However for fields applied parallel to the easy axis (small $\theta$) of a
mesoscopic wire, it was found that the switching field is one order of
magnitude smaller than the SW-model
predicts\cite{wern1,pignard,wegrowe,adey,chou,monzon}. To explain these low
switching fields two other switching mechanisms have been proposed: a
magnetization curling process and a domain-wall nucleation process.

The curling model assumes that the magnetization direction rotates in a plane
perpendicular to the anisotropy axis of the wire, effectively reducing the
longitudinal component of the magnetization and hence the magnitude of the
switching field\cite{brown,aharoni1,aharoni2}. For rectangular shaped strips,
the upper and lower bound of the magnitude of the switching field have been
calculated for a B-field applied parallel to the easy axis ($\theta=0^{\circ}$)
of the strip. For aspect ratios $d/h<4$, where $2d$ is width and $2h$ is the
height of the strip, these upper and lower bounds are the same. The magnitude
of switching field for a magnetic field applied parallel to the easy axis
($\theta=0^{\circ}$), as calculated by Aharoni, can then be written
as\cite{aharoni1}:

\begin{equation}
H_{c}^{curl}(0^\circ)=\frac{\pi}{2}M_s\frac{d_\circ^2}{d^2}\;, \label{aharoni}
\end{equation}

where $d_\circ = \sqrt{A/M_S^2}$ is the exchange length, $A$ being the exchange
constant and $M_S$ is the saturation magnetization. For permalloy we find that
$d_\circ \approx 12 nm$, using $A=~1\cdot 10^{-11}~J/m$ and $M_s=~860~kA/m$.
For a $100$ nm wide rectangular Py electrode and a field applied parallel to
the long (easy) axis, we would thus obtain a switching field of
$H_{c}^{curl}(0^\circ)=~91$ mT.

The other mechanism assumes that the switching of the magnetization is mediated
by the nucleation of a domain-wall \cite{smith,wern2}. A domain-wall is
nucleated (annihilated) when the cost of exchange energy associated with the
domain wall is lower (higher) than the gain in magnetostatic energy upon
increasing the external field. Once it is nucleated it sweeps through the
material, thereby lowering the total magneto static energy. This mechanism has
been conf\mbox{}irmed experimentally by Lorentz micrography by
Otani\cite{otani}. Recent MFM studies of $1 \mu m$ wide iron and permalloy
wires seem to indicate that in these wires a multi-domain structure is formed
during the reversal process\cite{kent,nitta}. However an analytical expression
of the magnitude of the nucleation field cannot easily be given, as one has to
numerically solve the time dependent Landau-Lifschitz equations for each value
of the applied magnetic field.

\subsection{The AMR behavior of rectangular Py electrodes}

\begin{figure}[htb]
\centerline{\psfig{figure=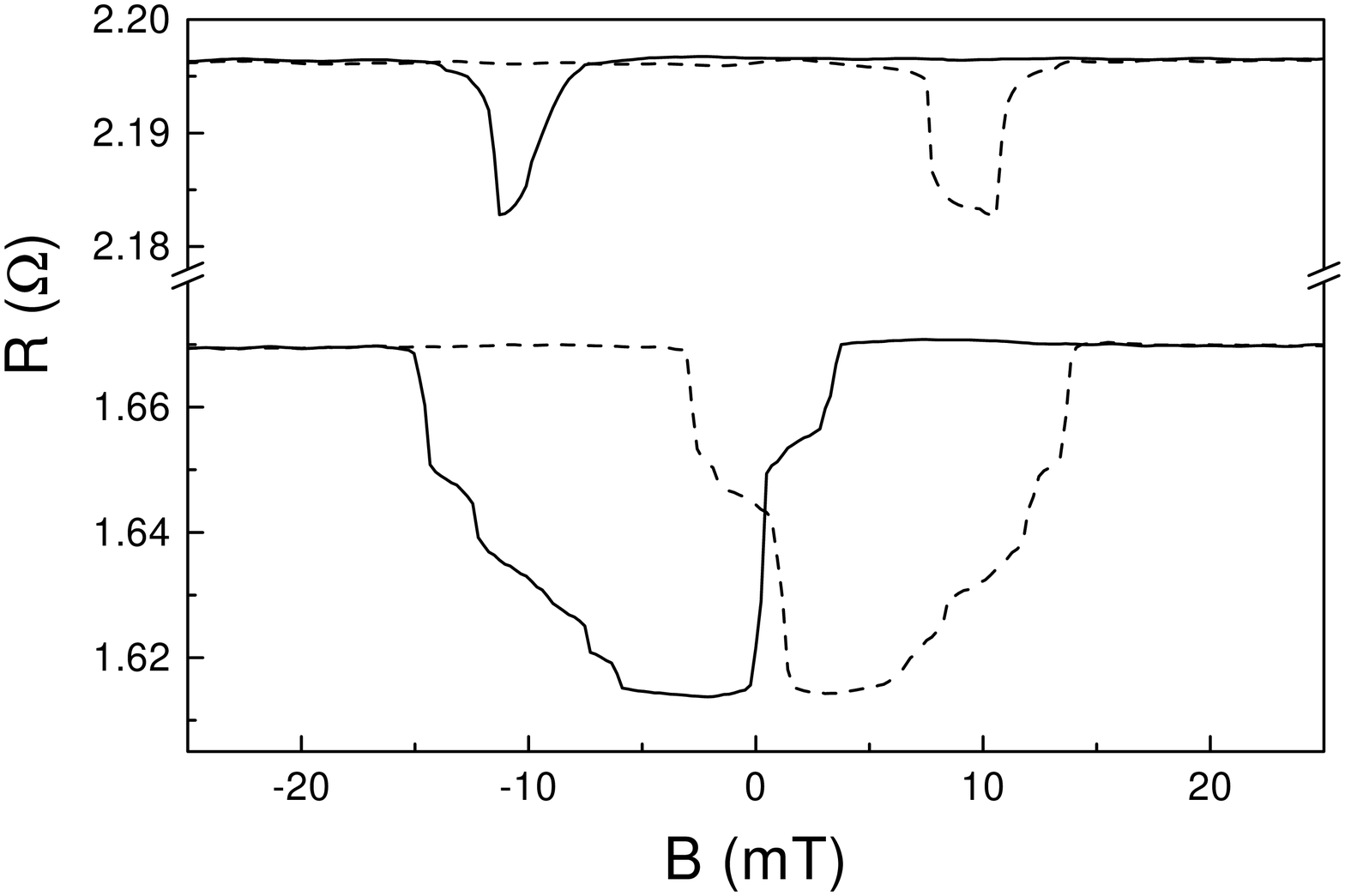,width=86mm}} \caption{Anomalous magneto
resistance (AMR) behavior at $T=4.2K$ of two rectangular Py electrodes with
dimensions $2.0 \times 0.8\mu m^2$ (bottom) and $2.0 \times 0.5\mu m^2$ (top).
The solid(dotted) curve corresponds with a negative (positive) sweep direction
of the B-field, which is applied parallel to the long (easy) axis of the Py
electrode.} \label{amrstrips}
\end{figure}

The AMR behavior of the $150$, $500$ and $800$ nm wide rectangular Py
electrodes used in the spin valve samples, as shown in fig. \ref{sample}, was
measured four terminal by using contacts $1$ and $4$ as current contacts and
$2$ and $3$ as voltage contacts. In fig. \ref{amrstrips} the magneto resistance
behavior at $T=~4.2$ K of the $2.0 \times 0.8~\mu m^2$ (bottom curves) and $2.0
\times 0.5~\mu m^2$ (top curves) sized Py electrodes is shown, where the
magnetic field is applied parallel to the long axis of the Py electrode
($\theta=0^{\circ}$).

Coming from a negative B-field the $2.0 \times 0.8\mu m^2$ electrode already
has a change in the resistance before the magnetic field reaches zero. After
this first drop in the resistance at $-3$ mT, a broad step like transition
range is observed up to $+15$ mT, which indicates that the Py strip breaks up
in a multiple domain structure. The amplitude of the AMR signal \cite{domain}
is about $3.3\%$ of the total resistance, which is a commonly reported value in
literature\cite{coehoorn}. The $2.0 \times 0.5\mu m^2$ Py electrode shows a
more 'ideal' switching behavior, showing only a resistance change after the
magnetic field has crossed zero and showing a much narrower transition range
from $7$ to $14$ mT. However, the amplitude of the resistance dip has changed
to $0.7 \%$. Taking the minimum of the resistance dip as the switching field we
find a value of $10$ mT, which is much below the SW switching field of $\mu_0
H_{0}^{sw}=~540 mT$. Applying eq. \ref{aharoni} to calculate the curling
switching field is not allowed, as the ratio $d/h$ of this electrode is bigger
than 4 ($d/h=~13$).

For the narrowest strip with a width of $150$ nm (see fig. \ref{sample}) we do
not observe any magneto resistance signal in parallel field, which is an
indication that this electrode behaves as a single domain or reverses its
magnetization by means of a fast domain wall sweep.

\subsection{Magneto resistance behavior of the Py/Cu contacts}

A possible formation of a domain structure in the Py electrodes is important
for a spin valve measurement, since the spin flip length of Py is very short
($\lambda\approx 5$ nm, see \ref{dub1}) as compared to the domain size. In case
of domain formation the magnetization direction of the injecting and detecting
electrodes would be determined by the local domain(s) present at the Py/Cu
contact area which could a have different magnetic switching behavior as the
entire Py electrode.

Therefore we have locally measured the magneto resistance at the Py/Cu contact
area, which we will call the "contact" magneto resistance. For example the
"contact" magneto resistance of the $14.0 \times 0.15~\mu m^2$ Py electrode can
be measured by sending current (see fig. \ref{sample}) from contact $6$ to $8$
and measuring the voltage with contacts $5$ and $7$. Note that in this geometry
one is not sensitive for a spin valve signal, as only \emph{one} Py electrode
is used in the measurement.

\begin{figure}[b]
\centerline{\psfig{figure=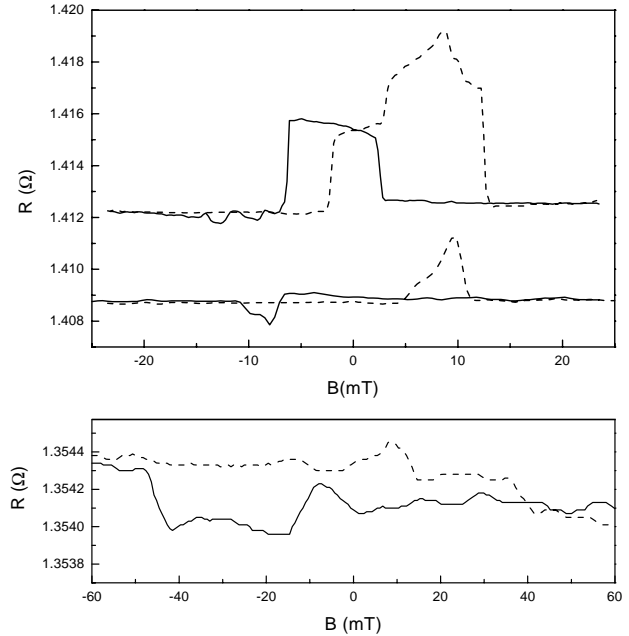,width=86mm}} \caption{'Contact'
magneto resistance behavior at $T=~4.2$ K of the Py/Cu contact area for Py
electrodes with dimensions $2.0 \times 0.8~\mu m^2$ (top), $2.0 \times 0.5~\mu
m^2$ (middle) and $14.0 \times 0.15~\mu m^2$(bottom). The solid(dotted) curve
corresponds with a negative (positive) sweep direction of the B-field, which is
applied parallel to the long (easy) axis of the Py electrode.}
\label{amrcontacts}
\end{figure}

Figure \ref{amrcontacts} shows the "contact" magneto resistance behavior at
$T=~4.2 K$ of three rectangular Py electrodes with dimensions: $2.0 \times
0.8~\mu m^2$, $2.0 \times 0.5~\mu m^2$ and $14.0 \times 0.15~\mu m^2$.

The "contact" magneto resistance of the $500$ and $800$ nm wide electrodes show
a similar magneto resistance behavior as the magneto resistance plots of the
entire strips shown in fig. \ref{amrstrips}, except that there seems to be more
asymmetry. For the $500$ nm wide electrode a 'positive' peak is shown in the
positive sweep direction and a 'negative' peak in the negative sweep direction.
This indicates that the magnetization reversal process is different for a
positive and negative magnetic field sweep, resulting in different domain
structures at the Py/Cu contact. However, it is important to note that
amplitude of the 'contact' magneto resistance can be as high as $7 m\Omega$ for
the $800$ nm wide Py electrode. This magnitude is large as compared to the
amplitude of the spin valve effect, as we will show in the next section.

For the $150$ nm wide Py electrode( bottom curve) a "contact" magneto
resistance behavior is observed, which appearance resembles much of a Hall
signal, showing a difference in resistance at large negative and positive
magnetic fields. A Hall voltage perpendicular to the substrate surface
(z-direction) can be expected, as Py electrode is etched prior to the Cu
deposition, causing the Cu wire to be a little bit 'sunk' into the Py
electrode. Changing one voltage probe from contact $7$ to contact $9$ (at the
other side of the Py/Cu contact area, see fig. \ref{sample}) produces the same
signal. Also the signal amplitude of $0.3~m\Omega$ lies in the range of a Hall
signal, which would have a maximum of $1~m\Omega$, using a Cu Hall resistance
of $1~m\Omega/T$ for a $50$ nm thick film and a maximal obtainable magnetic
field change upon magnetization reversal of about $1$ T. When we take the
position of the Hall step as the switching field at $42$ mT, we find a good
agreement with the curling switching field $H_{c}^{curl}(0^\circ)=~40$ mT for a
width $2d=~150$ nm.

\section{The Spin valve Effect}

Two different measurement geometries are used to measure the spin valve effect
in our device structure, the so called 'conventional' geometry and 'non-local
geometry. In the conventional measurement geometry the current is sent from
contact $1$ to $7$ and the signal $R=V/I$ is measured between contacts $4$ and
$9$, see fig. \ref{sample}. In the non-local measurement geometry the current
is sent from contact $1$ to $5$ and the signal $R=V/I$ is measured between
contacts $6$ and $9$, see also fig. \ref{diagram}a. This technique is similar
to the "potentiometric" method of Johnson used in ref. \ref{john1}. The
difference between the two measurement geometries is that the conventional
geometry suffers from a relatively large background resistance as compared to
the spin valve resistance. The bad news is that this background resistance
includes also small parts of the Py electrodes underneath the vertical Cu wires
of the cross and the Py/Cu interface itself, which give rise to the "contact"
magneto resistance as was described in the previous section. Experimental
measurements show that the spin valve signal can be completely dominated by the
"contact" magneto resistance of the Py eletrodes.

\subsection{Spin valve measurents}

\begin{figure}[htb] \centerline{\psfig{figure=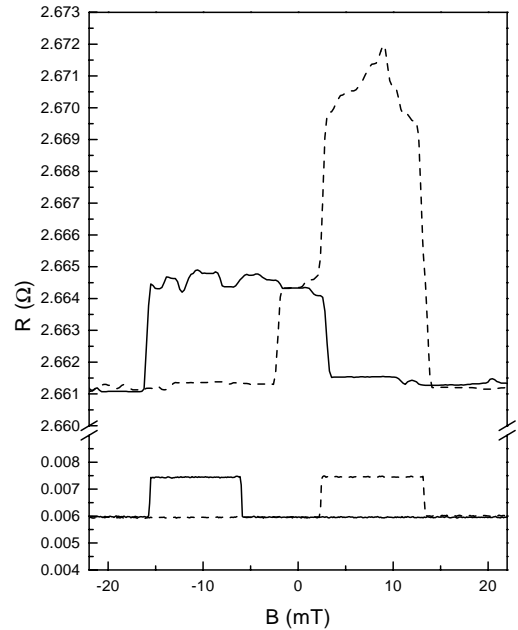,width=70mm}}
\caption{The spin valve effect using a conventional measurement geometry (top
curve) at $T= 4.2$ K and non-local measurement geometry (bottom curve), with a
Py electrode spacing $L= 250$ nm. The sizes of the Py electrodes are $2 \times
0.8\mu m^2$ (Py1) and $14 \mu m \times 0.5\mu m^2$ (Py2). The B-field is
applied parallel to the long (easy) axis of the Py electrodes. The
solid(dotted) curve corresponds with a negative (positive) sweep direction of
the B-field.} \label{spinbig}
\end{figure}

\begin{figure}[hbt] \centerline{\psfig{figure=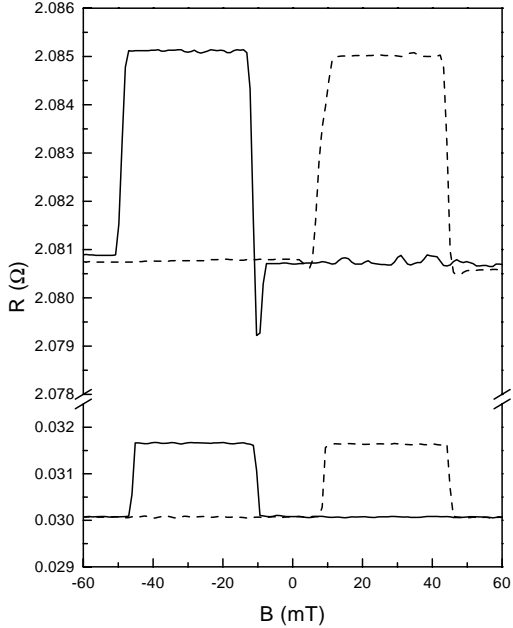,width=70mm}}
\caption{The spin valve effect using a conventional measurement geometry (top
curve) at $T= 4.2$ K and non-local measurement geometry (bottom curve), with a
Py electrode spacing $L= 250$ nm. The sizes of the Py electrodes are $2 \times
0.5\mu m^2$ (Py1) and $14 \mu m \times 0.15\mu m^2 (Py2)$. The B-field is
applied parallel to the long (easy) axis of the Py electrodes. The
solid(dotted) curve corresponds with a negative (positive) sweep direction of
the B-field.} \label{spinsmall}
\end{figure}

The measurements were performed by standard ac-lock-in-techniques, using
current magnitudes of $100~\mu A$. The spin valve signals of two different
samples (of the same batch) with a separation distance of $L=~250$ nm are shown
in fig. \ref{spinbig} and \ref{spinsmall}. The first sample, see fig.
\ref{spinbig} had a current injector Py electrode of size $2 \times 0.8~\mu
m^2$, whereas the detector electrode had a size of $12 \times 0.5~\mu m^2$. The
second sample, see fig. \ref{spinsmall} had narrower Py electrodes of $2 \times
0.5~\mu m^2$ and $14 \times 0.15\mu m^2$. This difference in width of the Py
electrodes can be observed in the increased switching fields of the second
sample.

From fig. \ref{spinbig} (top curves) we can see that the total magneto
resistance signal in in the conventional geometry is about $11 m\Omega$. The
amplitude of the spin valve signal, measured in the non local geometry, is
shown in the botton curve of fig. \ref{spinbig}. Sweeping the magnetic field
from negative to positive field, an increase in the resistance is observed,
when the magnetization of Py1 flips at $3$ mT, resulting in an anti-parallel
magnetization configuration. When the magnetization of Py2 flips at $14$ mT,
the magnetizations are parallel again, but now point in the opposite direction.
The magnitude of the spin valve signal measured in the non local geometry is
$1.5 m\Omega$ (at $4.2$ K), much lower that the magneto resistance signal of
$11 m\Omega$ measured in the conventional geometry. We therefore conclude that
the 'contact' magneto resistance of the $800$ and $500$ nm wide Py electrodes
are completely dominating the magneto resistance signal in a conventional
measurement geometry, making it impossible to detect a spin valve signal.

For the sample with Py electrodes of sizes $2 \times 0.5\mu m^2$ and $14 \times
0.15\mu m^2$ a spin valve signal can be observed in the conventional geometry.
This is shown in the top curve of fig. \ref{spinsmall}. A small magneto
resistance dip around $10$ mT can be observed in this sample upon switching
from parallel to the anti-parallel magnetization configuration. The position of
this peak in the magnetic field sweep and the amplitude of $2 m\Omega$
correspond to the 'contact' magneto resistance behavior of the $2 \times 0.5\mu
m^2$ Py electrode, see fig. \ref{amrcontacts}. However, after the magnetization
of this Py electrode has switched, we do observe a resistance 'plateau' up to a
magnetic field of $45$ mT, where the second $150$ nm wide Py electrode
switches. The magnitude of the spin valve effect measured in the conventional
geometry is about $4.1 m\Omega$. This is more than $2$ times bigger than the
magnitude of the spin signal of $1.6 m\Omega$ (at $T=~4.2$ K), measured in a
'non-local geometry, as shown in the bottom curve of fig. \ref{spinsmall} (see
also ref. \ref{jedema}). Calculations show (see \ref{filip}, eq. 3) that the
magnitude of the spin valve signal measured in a conventional geometry should
be twice the magnitude of the spin valve signal measured in the non local
geometry. At this moment we do not clearly understand why the measured ratio of
the two spin signals is slightly larger than $2$ (factor of $2.5$).

\subsection{Dependence on Py electrode spacing}

\begin{figure}[b]
\centerline{\psfig{figure=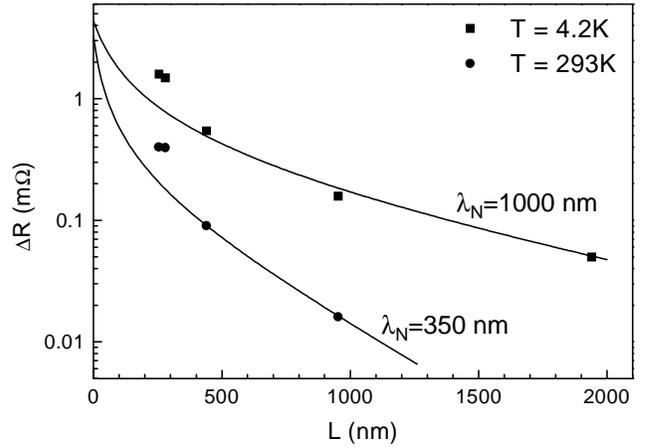,width=86mm}} \caption{Dependence of
the magnitude of the spin signal $\Delta R$ on the Py electrode distance $L$,
measured in the non local geometry. The solid squares represent data taken at
$T=~4.2$ K, the solid circles represent data taken at RT. The solid lines
represent the best fits based on equation \ref{Rspinfull}.} \label{ldepent}
\end{figure}

A reduction of the magnitude of spin signal $\Delta R$ is observed with
increased electrode spacing $L$ , as shown in fig. \ref{ldepent}. By fitting
the data to eq. \ref{Rspinfull} we have obtained $\lambda_{N}$ in the Cu wire.
From the best fits we find a value of $1~\mu m$ at $T=~4.2$ K, and $350$ nm at
RT. These values are compatible with those reported in literature, where $450$
nm is obtained for Cu in GMR measurements at $4.2$ K\cite{yang}. However one
should be careful to make a straightforward comparison between the GMR results
and ours. In the thin films we use, the elastic mean free path $l_e$ of the
electrons is limited by surface scattering, causing the conductivity of the Cu
to be smaller than in GMR layers.

We can calculate the spin flip time $\tau_{N}$ in the Cu wire, using a Fermi
velocity $1.59\cdot 10^6$ m/s \cite{ashcroft}. At $T=~4.2$ K we find
$\tau_{N}=~42$ ps, while at RT $\tau_{N}=~11$ ps. Comparing the spin flip time
to the elastic scattering time $\tau_e=~2.9\cdot 10^{-14}$ s at $T=~4.2$ K, we
find that on average the spin is flipped after about $10^3$ elastic scattering
events in the Cu wire.

In principle the fits of fig. \ref{ldepent} also yield the spin polarization
$\alpha_F$ and the spin flip length $\lambda_{F}$ of the Py electrodes.
However, the values of $\alpha_F$ and $\lambda_{sf}^F$ cannot be determined
separately, as in the relevant limit ($M~>>~1$) which applies to our experiment
($12~<<~M<<~26$), the spin signal $\Delta R$ is proportional to the product
$\alpha_F\lambda_{F}$. From the fits we find that $\alpha_F\lambda_{F}=~1.2$ nm
at 4.2 K and $\alpha_F\lambda_{F}=~0.5$ nm at RT. Taking, from literature
\cite{dub1}, a spin flip length in the Py electrode of $\lambda_{F}=~5$ nm (at
4.2 K), a bulk current polarization of $22~\%$ in the Py electrodes at $T=~4.2$
K is obtained: $\alpha_F=~0.22$ . These values are in the same range as the
results obtained from the analysis of the GMR
effect.\cite{gijs1,ans1,dub1,yang}. However, the current polarization P
\emph{at the interface} of the current injecting Py electrode is much lower.
Using eq.\ref{polint1}, a polarization P for the samples with the smallest Py
electrode spacing of $L=~250$ nm at $T=4.2$ K is found to be only $2 \%$:
$P=0.02$. The reason for this reduction is caused by the unfavorable ratio of
the 'small' spin dependent resistance ($\lambda_F/\sigma_F$) and the 'large'
spin independent resistance ($L/\sigma_N$), which applies even in the absence
of spin flip scattering events in the normal (Cu) metal.

\section{conclusions}

We have demonstrated spin injection and accumulation in a mesoscopic spin
valve. We have shown that in conventional measurement geometry the magneto
resistance effects of the injecting and detecting contacts can be much larger
than the spin valve effect. These contact effects can be used to monitor the
magnetization reversal process of the spin injecting and detecting contacts. In
a non-local measurement geometry we can completely isolate the spin valve
effect, as was reported earlier in ref. \ref{jedema}. Using this geometry we
find a spin flip length in Cu of around $1~\mu m$ at $T=~4.2$ K and $350$ nm at
RT. For the smallest Py electrode spacing, the magnitude of the spin signal and
the current polarization P in the Cu wire are limited by the unfavorable ratio
of the spin independent resistance of the Cu strips ($L/\sigma_{N}$) and the
spin dependent resistance of the Py ferromagnet ($\lambda_{sf}^F/\sigma_{F}$).

The authors wish to thank H. Boeve, J. Das and J. de Boeck at IMEC (Belgium)
for support in sample fabrication and the Stichting Fundamenteel Onderzoek der
Materie for financial support.


\begin{references}
\bibitem{domain} We expect the largest part of the magneto resistance signal to
originate from the AMR effect. However we cannot exclude other (smaller)
contributions, such as a possible domain wall resistance.
\end{references}
\end{document}